\documentclass{iopart}

\usepackage{epsfig}
\usepackage{graphicx}
\usepackage{bbm}

\usepackage{epstopdf}
\newcommand{\be}{\begin{equation}}
\newcommand{\ee}{\end{equation}}

\newcommand{\cMpc}{{\mathrm{Mpc}}^{-3}}

\begin{document}

\title{Mock Catalogs for UHECR Studies}

\author{A. Berlind$^1$, N. Busca$^2$, G. R. Farrar$^3$, J. P. Roberts$^3$}

\address{
$^1$Department of Physics and Astronomy, Vanderbilt University,\\
Nashville, Tennessee 37240, USA \\
$^2$Laboratoire Astroparticule et Cosmologie (APC), UniversitŽ Paris 7/CNRS\\
10, rue Alice Domon et LŽonie Duquet, 75205 Paris Cedex 13, France\\
$^3$Center for Cosmology and Particle Physics, Department of Physics\\
New York University, NY, NY 10003, USA\\
}


\begin{abstract}
We provide realistic mock-catalogs of cosmic rays above 40 EeV, for a pure proton composition, assuming their sources are a random subset of ordinary galaxies in a simulated, volume-limited survey, for various choices of source density: $10^{-3.5} ~\cMpc$, $10^{-4.0} ~\cMpc$ and $10^{-4.5} ~\cMpc$. The spectrum at the source is taken to be $E^{-2.3}$ and the effects of cosmological redshifting as well as photo-pion and $e^+$ $e^-$ energy losses are included.\end{abstract}

In order to develop and test analysis tools for ultrahigh energy cosmic ray (UHECR) correlation studies it is necessary to have mock catalogs of UHECRs under various realistic scenarios for the sources and magnetic deflections. There is no need at present to postulate a ``new physics'' origin for UHECRs, so the sources of UHECRs will be associated with the distribution of matter in the universe. The most popular types of source candidates reside in ordinary galaxies - for instance, AGNs and GRBs - and to first approximation these galaxies are a random subset of all galaxies. Here we provide mock UHECR catalogs assuming the sources are randomly selected galaxies. In this case, the key characteristic of a source population is its number density. The small number of doublets in the Auger events above 57 EeV leads to the lower bound of 61 on the number of sources producing them \cite{Auger2008}; this in turn leads to a lower bound on the source density of about $3\times10^{-5} ~\cMpc$ \cite{farrarGruzinov}. Simulations of the spectrum suggest a density of $10^{-4.0} ~\cMpc$ \cite{Takami}. Therefore, we provide catalogs for the source density hypotheses $10^{-3.5} ~\cMpc$, $10^{-4.0} ~\cMpc$ and $10^{-4.5} ~\cMpc$.

\section{Mock Galaxy Catalogs}
To have a realistic treatment of CRs in the GZK transition region from 40-60 EeV, the galaxy catalog used to produce the UHECR mocks must be volume-limited to a much larger depth than is available in present-day ``all sky'' galaxy surveys, the best of which are 2MRS\cite{2MRS} and PSCz\cite{PSCz}. In particular, the galaxy catalog from which the source catalog is derived must be considerably denser than $10^{-3.5} ~\cMpc$ to simulate a scenario in which the source catalog has that density, and yet not every galaxy hosts a UHECR source. Therefore, we make use of the ``LasDamas'' mock galaxy catalogs \cite{McBride} which are created using LCDM simulations ($\Omega_M = 0.25$, $\Omega_{DE} = 0.75$, $H_0 = 70~\mathrm{km}/\mathrm{s}/\mathrm{Mpc}$, $\sigma_8 = 0.8$ and $n_s = 1$) combined with an algorithm for populating the DM halos with galaxies, that is tuned to reproduce the clustering of galaxies as measured in the SDSS\cite{SDSS}.

We start with two LasDamas ``Carmen'' catalogs which are volume-limited for $M_r < -21$ (includes all galaxies with an absolute r-band magnitude of -21 and brighter) to a redshift $z = 0.165$, and have a number density of $4\times10^{-4.0} ~\cMpc$ or $\log(n) = -3.41$ . They are:

\begin{itemize}
\item Carmen02\_Mr21\_Master.dat - 522769 galaxies
\item Carmen27\_Mr21\_Master.dat - 520277 galaxies
\end{itemize}

\begin{figure}
\begin{center}
\scalebox{0.35}{\includegraphics{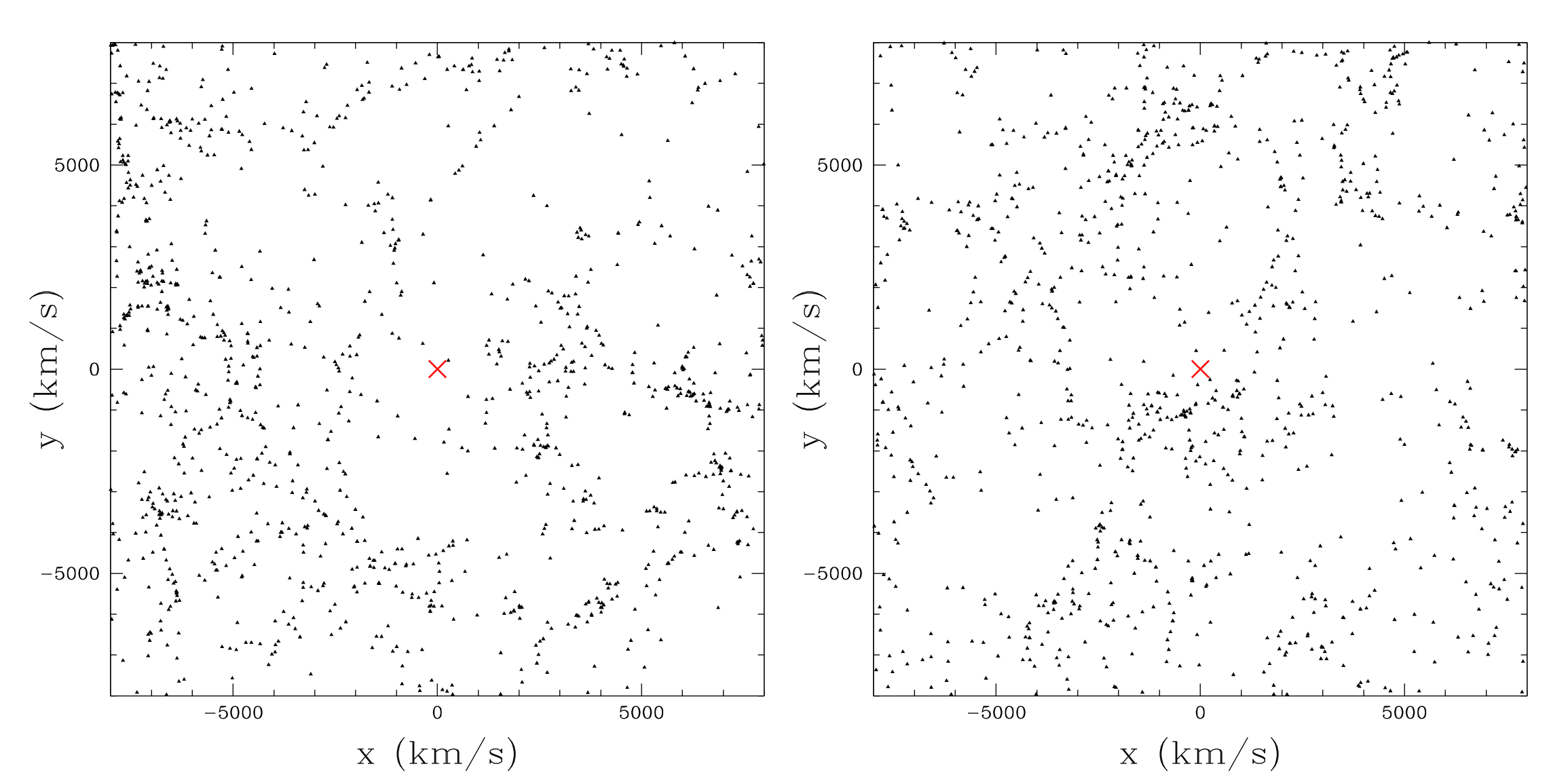}}
\end{center}
\caption{Local density of galaxies in Carmen02\_Mr21\_Master (left) and Carmen27\_Mr21\_Master (right) in a slice of thickness 4000 km/s centered on the origin.\label{fig:localStructure}}
\end{figure}

The local structure in the two catalogs can be seen in Fig.~\ref{fig:localStructure}. The galaxies in each file are ordered by increasing redshift and each has a unique ID with the format mgC$\langle$catID$\rangle$\_$\langle$N$\rangle$ where catID is either 02 or 27. N is a unique integer that corresponds to the galaxy position in the redshift ordered list.

We randomly subsample each catalog to obtain sets of catalogs at lower densities. We create one catalog with log(n)=-3.5 for each of Carmen02 and Carmen27:
\begin{itemize}
\item mockgalsCarmen02,3.5.dat
\item mockgalsCarmen27,3.5.dat
\end{itemize}

We create 3 independent mock UHECR-source catalogs with log(n) = -4 for each of Carmen02 and Carmen27:
\begin{itemize}
\item mockgalsCarmen02,4.0,R$\langle$i$\rangle$.dat where i=1,2,3
\item mockgalsCarmen27,4.0,R$\langle$i$\rangle$.dat where i=1,2,3
\end{itemize}

We create 3 independent mock UHECR-source catalogs with log(n) = -4.5 for each of Carmen02 and Carmen27:
\begin{itemize}
\item mockgalsCarmen02,4.5,R$\langle$i$\rangle$.dat where i=1,2,3
\item mockgalsCarmen27,4.5,R$\langle$i$\rangle$.dat where i=1,2,3
\end{itemize}

In all of these source catalogs, the file format is {\tt RA Dec	 z id}. The ID field gives the ID of the galaxy to allow comparison with the relevant master catalog.

\begin{figure}
\begin{center}
\scalebox{0.5}{\includegraphics{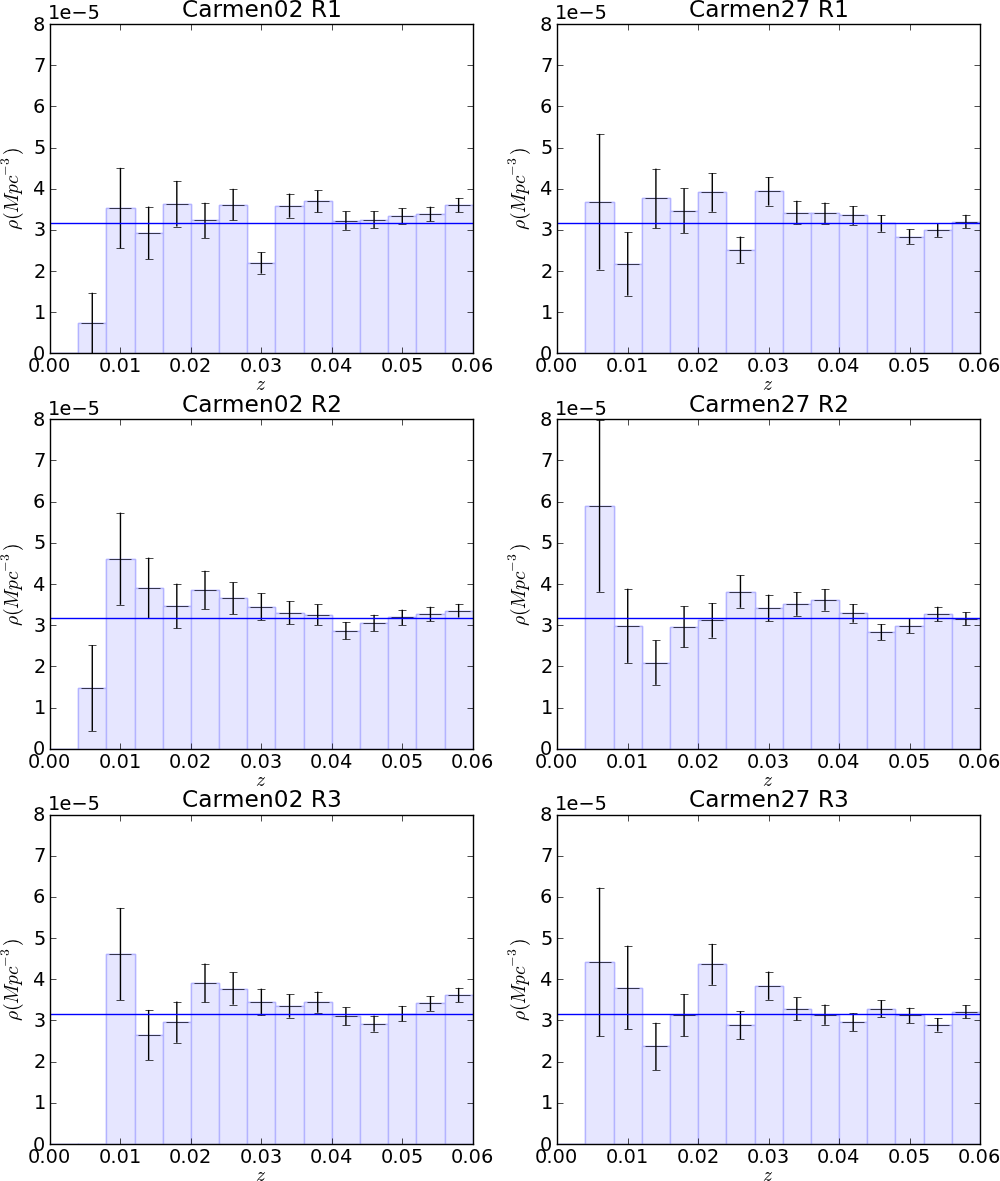}}
\end{center}
\caption{The density of galaxies per redshift bin in each of the 6 Carmen catalogs with $\rho=10^{-4.5}\cMpc$.\label{fig:DensityHist}}
\end{figure}

The density of galaxies per redshift bin is shown in Fig.~\ref{fig:DensityHist} for redshifts within $z=0.06$. The error bars on the histogram show the Poisson error for the number of galaxies in the bin. The lack of any galaxies within $z=0.0045$ for Carmen 02 and within $z=0.0025$ for Carmen 27 is a feature of the master catalogs. Carmen02\_Mr21\_Master.dat contains no galaxies within $z=0.0045$, Carmen27\_Mr21\_Master.dat has 18 galaxies within $z=0.0045$, but none within $z=0.0025$. The lack of very nearby galaxies is intentional, to simulate the underdense region around the Milky Way.

\begin{figure}
\begin{center}
\scalebox{0.5}{\includegraphics{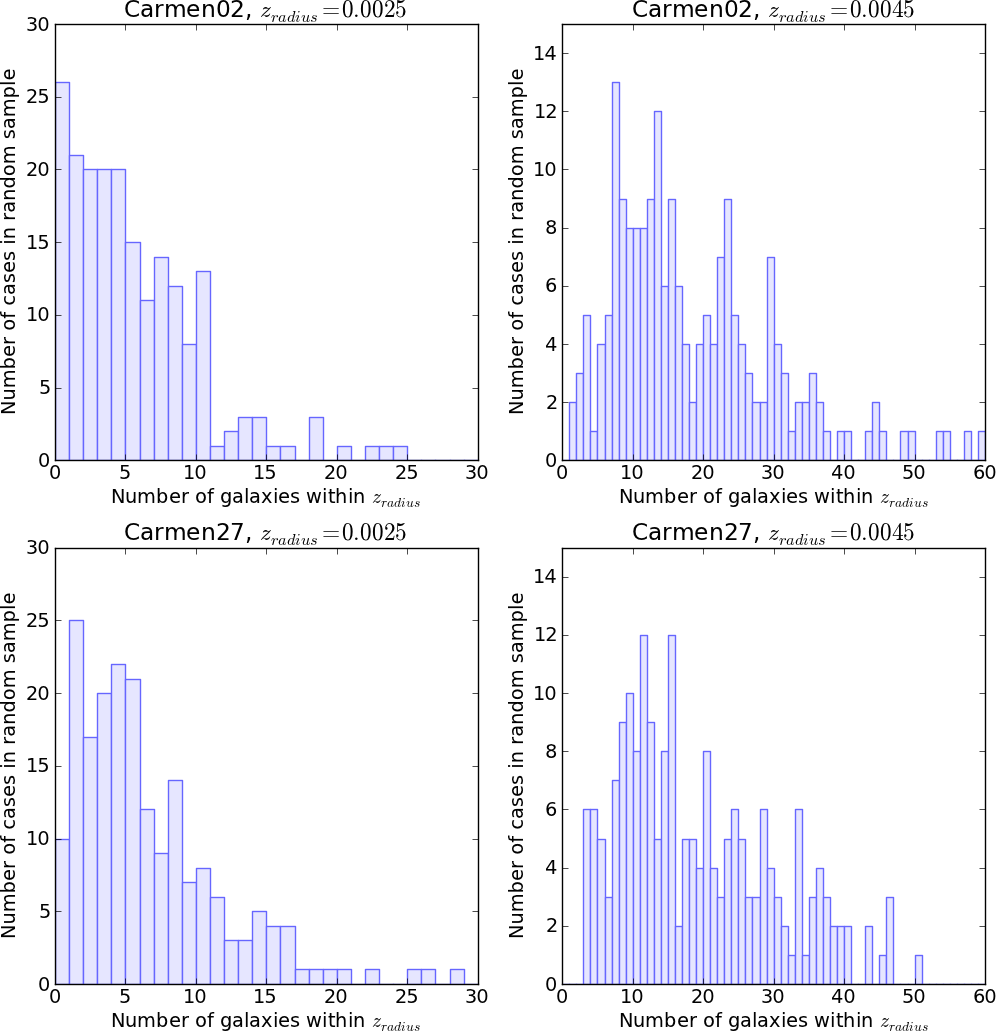}}
\end{center}
\caption{The number of galaxies within $z_{radius}$ of a candidate central galaxy, chosen at random from the master catalogs of Carmen 02 and Carmen 27. Each plot shows the results of 200 randomly chosen central galaxies. \label{fig:randomSpread}}
\end{figure}

Fig.~\ref{fig:randomSpread} shows the result of taking 200 random galaxies from the full catalog (Carmen02\_Mr21\_Master.dat or Carmen27\_Mr21\_Master.dat) as central reference points, and counting the number of galaxies within a spherical volume with radius $z_{radius}=0.0025$ or $z_{radius}=0.0045$ around each central point. This shows that an underdense region within $z=0.0045$ is not a generic feature for a randomly selected central galaxy.  

\section{Mock Cosmic Rays}

\begin{figure}
\begin{center}
\scalebox{0.4}{\includegraphics{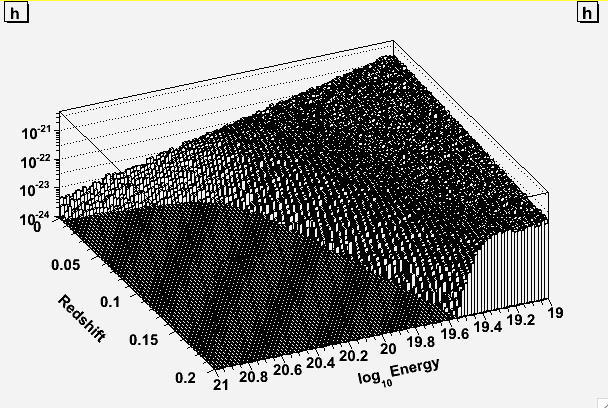}}
\end{center}
\caption{Distribution of energies at Earth, as a function of redshift of source.\label{fig:CRPropagation}}
\end{figure}

From these mock-source catalogs, we create mock UHECR catalogs from a given mock-source catalog by choosing a galaxy at random, with a probability $p_i = w_i/\Sigma_j w_j$, with $w_i=r_i^{-2}$, where $r_i$ is the distance of the $i$th galaxy. (Distance, not redshift, is used here because the effects of peculiar motion can lead to up to 20 Mpc shifts in the distance with respect to that calculated from the Hubble flow.) A CR from this galaxy is then assigned an energy according to an injection energy spectrum and propagated to Earth. Energy loss from redshifting as well as photo-pion and e+e- production is included. In the simulations presented here, the CR is assumed to be a proton; the injection spectrum is proportional to $E^{-2.3}$ and ranges from 40 to 1000 EeV. If the energy after propagation is greater than 40 EeV, this proton is kept as an event. This procedure of source sampling and propagation is repeated until the desired number of events have been kept. Fig. \ref{fig:CRPropagation} shows the final energy distribution as a function of redshift (actually, distance expressed as an equivalent redshift without peculiar motion). The resulting catalogs are:

\begin{itemize}
\item mockgalsCarmen02,3.5,CRs.dat
\item mockgalsCarmen27,3.5,CRs.dat
\item mockgalsCarmen02,4.0,R$\langle$i$\rangle$,CRs.dat, i=1,2,3
\item mockgalsCarmen27,4.0,R$\langle$i$\rangle$,CRs.dat, i=1,2,3
\item mockgalsCarmen02,4.5,R$\langle$i$\rangle$,CRs.dat, i=1,2,3
\item mockgalsCarmen27,4.5,R$\langle$i$\rangle$,CRs.dat, i=1,2,3
\end{itemize}

Each cosmic ray catalog includes 100,000 mock cosmic rays, generated using the corresponding mock-galaxy catalog. The file format is:

{\tt CRId	 E(EeV)	 srcID	 srcZ	 srcRA	 srcDec	 InjectionE	 Ra1Deg ...	Ra8Deg	 Dec8Deg}

\begin{itemize}
\item {\tt CRId} is an integer id
\item {\tt E} is the energy of the CR at earth
\item {\tt srcID} is the ID of the source galaxy
\item {\tt srcZ}, {\tt srcRA} and {\tt srcDec} are the redshift, RA and Dec of the source galaxy
\item {\tt InjectionE} is the energy of the cosmic ray at the source (in eV)
\item{\tt Ra$\langle$j$\rangle$Deg}, {\tt Dec$\langle$j$\rangle$Deg} give smeared arrival directions for a smearing of $j$ degrees.
\end{itemize}

The columns headed {\tt Ra$\langle$j$\rangle$Deg}, {\tt Dec$\langle$j$\rangle$Deg} provide a range of arrival directions for the cosmic ray assuming a Gaussian deflection of the arrival direction:
\begin{equation}
\frac{dP(\theta)}{d\Omega}=\frac{1}{2\pi\sigma^2}\exp\left(-\frac{\theta^2}{2\sigma^2}\right).
\end{equation}

For this distribution a cone of radius $1.5\sigma$ contains 68\% of the smeared events. This angle is designated $\sigma_{68}$. {\tt Ra$\langle$j$\rangle$Deg}, {\tt Dec$\langle$j$\rangle$Deg} in the header columns refer to $\sigma_{68}$ to label the arrival directions with increasing angular deflections.

The full files can be downloaded from the project web page \cite{LasLlamas}\footnote{Prior to 2011 there were a set of mock galaxy catalogs and a corresponding set of mock cosmic rays. In these files there were 6 mock galaxy sets - one for each density and for each of Carmen 02 and Carmen 27. The files for these superseded mock catalogs are also available from the mock catalog web page for comparison \cite{LasLlamas}. The galaxy files in these legacy mock catalogs have the file format {\tt Ra Dec cz}. The cosmic ray mock catalogs have the format {\tt Ra Dec E(EeV) id\_src z\_src N\_src}, where the energy is the energy at $z=0$, and {\tt N\_src} is the number of cosmic rays in the set that originated from the same source galaxy.}.

\section{Attribution}
For proper attribution, we ask:
\begin{enumerate}
\item Any use must credit the mock UHECR catalog project and cite this paper and the LasDamas paper about the galaxy catalog from which it is derived \cite{McBride}.
\item If used in a publication, inform us about your work before submission.
\item Let us know that you downloaded the files, by sending an email to gf25@nyu.edu, so we can alert you if there are any problems discovered and about improvements implemented.
\end{enumerate}

By downloading the mocks, you implicitly agree to these terms.

\section{Future Work}
The Carmen mock catalogs will be extended in the near future to include mixed composition cosmic rays. The work will also be extended to provide mock cosmic ray catalogs for the known local distribution of galaxies from 2MRS.

\section*{Acknowledgements}

NB, GRF and JPR acknowledge their membership in the Pierre Auger Collaboration and thank their colleagues for useful discussions.  The research of GRF and JPR has been supported in part by NSF-PHY-0701451, NSF-PHY-0900631 and NSF-PHY-0970075. AAB has been supported by Vanderbilt University and the Alfred P. Sloan Foundation.

\end{document}